\documentclass[12 pt]{article}

\begin{document}
\setlength{\baselineskip}{2\baselineskip}	
\title{Radial Quantization in Rotating Space-Times}
\author{Robert D. Bock \footnote{Present address: Propagation Research Associates, Inc., 1275 Kennestone Circle, Suite 100, Marietta, GA, 30066} \footnote{email: robert.bock@pra-corp.com} \\ Electro-Optical Systems Laboratory, \\ Georgia Institute of Technology, Atlanta, GA 30332}
\date{\today}

\maketitle

\begin{abstract}
\setlength{\baselineskip}{2\baselineskip}	
We examine the time discontinuity in rotating space-times for which the topology of time is $S^1.$  A kinematic restriction is enforced that requires the discontinuity to be an integral number of the periodicity of time.  Quantized radii emerge for which the associated tangential velocities are less than the speed of light.  Using the de Broglie relationship, we show that quantum theory may determine the periodicity of time.  A rotating Kerr-Newman black hole and a rigidly rotating disk of dust are also considered; we find that the quantized radii do not lie in the regions that possess CTCs.
\end{abstract}

                             
\section{\label{sec:introduction}INTRODUCTION}
There is a long-standing debate in the literature concerning rotating frames of reference and rotating distributions of matter in Einstein's theory of relativity.  Much of this discussion focuses on the paradoxes associated with the global properties of time.  Both the time discontinuity \cite{relrotbook} as well as the existence of closed timelike curves (CTCs) \cite{lobocrawford} in certain exact solutions of Einstein's field equations have garnered much attention.

The time discontinuity (or time lag) arises when one tries to establish standard simultaneity along a closed curve in a rotating coordinate system.  Upon traversing a complete circuit in such a frame of reference an observer discovers that a clock situated at the curve's orgin is not synchronized with itself.  This is often treated in the context of special relativity alone.  According to the traditional viewpoint (see, for example, Refs. \cite{dieks, weber,anandan,bergiaguidone,cranoretal,rizzitartaglia}) special relativity is valid in rotating frames of reference and the time discontinuity is only an apparent problem. This traditional approach maintains that multiple clock readings at a given event, depending on the chosen synchronization procedure, are indeed acceptable.  Furthermore, it is argued that the time gap is no more problematic than the discontinuity in time at the International Date Line or the coordinate discontinuity in angle at 2$\pi$.  On the other hand, many authors have questioned the validity of special relativity in rotating frames of reference and have attempted to modify Einstein's postulates for rotational motion.  For example, Klauber \cite{klauber2} and independently, Selleri \cite{selleri2} contend that the synchronization procedure cannot be chosen freely for the rotating frame and propose a unique (non-Einstein) synchronization along the circumference.  

Closed timelike curves are also the subject of much debate in Einstein's theory of relativity.  A CTC is a future directed timelike curve in the space-time manifold that runs smoothly back into itself.  As is well known \cite{nahin}, the existence of CTCs suggests that time travel is compatible with general relativity since an observer may evolve in time within the future light cone and return to an event that coincides with an earlier departure from that event.  A number of exact solutions of the Einstein field equations exhibit nontrivial CTCs, including a rapidly rotating infinite cylinder \cite{vanstockum, tipler}, the G\"{o}del universe \cite{godel}, a Kerr black hole \cite{carter}, and spinning cosmic strings \cite{deseretal,gott}.  While the G\"{o}del universe, the cosmic strings and the van Stockum cylinder all possess properties that may be deemed unphysical, the low angular momentum Kerr black hole is believed to possess physical relevance - it is the unique final state of gravitational collapse \cite{wald}.  Therefore, CTCs cannot be dismissed simply as mathematical curiosities.  Furthermore, the proliferation of new solutions that exhibit CTCs \cite{mallett,bonnor2,bonnor3,bonnorward,bicakpravda} suggests that their appearance in general relativity poses a critical problem to the foundations of physics \cite{bonnor}.

Hawking \cite{Hawking} has suggested that quantum effects prevent the emergence of CTCs.  In particular, he showed that divergences in the energy momentum tensor caused by vacuum polarization effects create singularities prior to the appearance of CTCs.  Based on these results Hawking proposed the chronology protection conjecture:  the laws of physics do not allow the appearance of CTCs.  Kim and Thorne \cite{kimthorne} have suggested otherwise, namely, that the divergences in the energy momentum tensor may be cut off by quantum gravitational effects.  Without a well-defined theory of quantum gravity this matter is still open to debate \cite{lobocrawford}.

In the following we show that the time discontinuity and the problem of CTCs may be intimately related, namely, that a solution to the former may resolve the latter.  Specifically, we show that if time possesses a topology $S^1$, then a kinematic condition may be imposed on the permitted time gaps, which results in a series of preferred quantized radii.  For the case of the rotating frame of reference, the tangential velocities associated with these preferred radii are less than the speed of light.  In the Kerr-Newman black hole and the rotating disk of dust, the preferred radii do not lie in the regions that possess CTCs.  

\section{\label{sec:rotframe}ROTATING FRAME OF REFERENCE}
Consider a Minkowski space-time with cylindrical coordinates $\{T,R,\Phi,Z\}$.  The line element is given by:
\begin{equation}
\label{flatlineelement}
ds^2=c^2dT^2-dR^2-R^2d\Phi^{2}-dZ^2.
\end{equation}
The coordinate transformation from the laboratory frame $\{T,R,\Phi,Z\}$ to the rotating frame $\{t,r,\phi,z\}$ is given by:
\begin{equation}
\label{coordtransformation}
T=t, R=r, \Phi=\phi+\omega t, Z=z, 
\end{equation}  
where $\omega$ is the angular velocity of the rotating system as observed from the laboratory frame.  Substituting (\ref{coordtransformation}) into (\ref{flatlineelement}) gives:
\begin{equation}
\label{rotlineelement}
ds^2=\gamma^{-2}c^{2}dt^{2} - 2c\beta rd\phi dt-dr^{2}-r^{2}d\phi^{2} -dz^{2},
\end{equation}
where $\beta=\omega r /c<1$ and $\gamma = (1-\beta^2)^{-1/2}$. Note that the condition $\beta<1$ is arbitrarily imposed on the coordinate transformation.

The time discontinuity appears when one tries to establish standard simultaneity along the circumference of a circle in the rotating coordinate system.   Consider two clocks, A and B, separated by the infinitesimal distance $rd\phi$ in the rotating frame.   In order to define standard simultaneity between the two (infinitesimally near) clocks the time on clock B must be adjusted by the amount \cite{landaulifshitz}:
\begin{equation}
\label{localtimelag}
c\Delta t = -\beta\gamma^{2}rd\phi.
\end{equation}  
The well-known expression for the time discontinuity is obtained by integrating around the entire circumference of the circle:
\begin{equation}
\label{timegap}
\Delta t = -\frac{2\pi\beta\gamma^{2}r}{c}.
\end{equation}
Thus, if one sends a light ray from a clock A around the entire circumference of the circle, establishing standard simultaneity along the way, then one discovers that the clock at A must be adjusted by the amount $\Delta t = -\frac{2\pi\beta\gamma^{2}r}{c}$ in order for it to be synchronized with itself.  In other words, one discovers a problem when clocks are synchronized in the usual Einstein manner (i.e., using light rays and no physical motion of the clocks) along the circumference of the circle.  While nearby clocks on an open curve can be synchronized by adjusting the readings of the various clocks according to Eq. (\ref{localtimelag}), this procedure cannot be extended globally since $\Delta t$ in Eq. (\ref{localtimelag}) is not a total differential in $r$ and $\phi$.  That is to say, the synchronization procedure is path dependent in the rotating frame of reference.  

If we consider, however, a space-time manifold for which time possesses a topology $S^1$, then a kinematic restriction can be placed on the motion that resolves the problem.  Therefore, let us assume that coordinate time is periodic with period $t^{\ast}(r)$ in the rotating system at radius $r$.  If one sends a light ray from the observer around the entire circumference of the circle then the time discontinuity is consistent with the topology of time if the following condition is satisfied: 
\begin{equation}
\label{quantcond1}
|\Delta t | =nt^{\ast}(r),
\end{equation}
where $n$ is an integer.  In other words, if we assume that time is a periodic phenomenon, then the observed time lag is not problematic if it is equal to an integral number of the temporal period.  The kinematic condition may also be understood in the following manner:  The infinitesimal Lorentz transformation of time to the rotating frame involves the angular coordinate $\phi$ so that the transformation, globally, endows time with a topology $S^1$.  If the time coordinate itself possesses a topology $S^1$, then the transformation must be defined so that the periods match (up to an integral number of the period).  

A space-time with periodic time is defined by the equivalence of the coordinates $t$ and $t+nt^{\ast}$; this equivalence may or may not be accompanied by circumnavigation.  Hence, the following identification of coordinates is consistent with the kinematic condition (\ref{quantcond1}):
\begin{eqnarray}
\{t,r,\phi,z\} &=&  \{t+t^{\ast},r,\phi,z\} \\  
\{t,r,\phi,z\} &=&  \{t+t^{\ast},r,\phi+2\pi,z\}. 
\end{eqnarray}
Since the former condition implies a quantization of mass that has not been observed in nature, we adopt the latter condition, which implies quantization of angular momentum, as the definition of the periodicity of time.  As a result, CTCs are introduced into the manifold because trajectories exist for which a future-directed timelike observer may return to an earlier time.  At first glance, it may seem that we have removed one problem by replacing it with another.  However, such a manifold may indeed represent the physical space-time of our experience as long as the temporal period is much greater than or much less than the observed temporal scales of physics.

We now consider the case $t^{\ast}(r)\equiv t^{\ast}_{0}$, where $t^{\ast}_{0}$ is a constant.  Substituting Eq. (\ref{timegap}) into Eq. (\ref{quantcond1}) gives:
\begin{equation}
\label{rotatingquantradii}
r=n^{1/2}\frac{l^{\ast}_{0}}{\sqrt{1+n\left(\frac{\omega l^{\ast}_{0}}{c}\right)^2}},
\end{equation}
where $l^{\ast 2}_{0}\equiv\frac{t_{0}^{\ast}c^2}{2\pi\omega}.$  Hence, radial quantization emerges in the rotating frame of reference.  Furthermore, the resulting tangential velocities associated with the above radii remain less than the speed of light for all values of $n$.  This is easiest to see in the limit $n\rightarrow\infty$ for which Eq. (\ref{rotatingquantradii}) gives $\omega r\rightarrow c$.  Thus, the kinematic condition (\ref{quantcond1}) also prevents causality violation from appearing in the rotating coordinate system.  Previously, this condition was imposed arbitrarily on the coordinate transformation to the rotating frame.        

The physical mechanism that is responsible for the topology of time remains to be determined.  We now show that quantum theory can serve as a guide.  We consider a mass $m_0$ at rest in the rotating system.  Let us assume that the periodicity of time, $\tau^{\ast}$, in the proper frame of the mass is given by the fundamental period of quantum theory:
\begin{equation}
\label{taustar}
\tau^{\star}=\frac{h}{m_0c^2}.
\end{equation}
In the rotating frame of reference the coordinate time corresponding to the proper time $\tau^{\ast}$ is:
\begin{equation}
\label{massperiodcoord}
t^{\ast} = \frac{\tau^{\ast}}{\sqrt{\left(1-\frac{\omega^2r^2}{c^2}\right)}}.
\end{equation}
Substituting Eqs. (\ref{taustar}) and (\ref{massperiodcoord}) into the kinematic condition (\ref{quantcond1}) gives:
\begin{equation}
\label{stationarystates}
\frac{m_0\omega r^2}{\sqrt{\left(1-\frac{\omega^2r^2}{c^2}\right)}}=n \hbar.
\end{equation}
The reader will readily note that Eq. (\ref{stationarystates}) is the quantum condition discovered by de Broglie \cite{debroglie} for the stationary states of a rotating electron.  Thus, the well-known quantum condition acquires a kinematic interpretation in the present work.  Time itself is periodic with period $\frac{h}{m_{0}c^{2}}$ in the rest frame of a particle of mass $m_{0}$.  A mass in a rotating frame of reference is restricted to occupy only those rotational states for which the time discontinuity is compatible with the period defined by $m_{0}$.    

Generalizing this result to an arbitrary mass distribution, we postulate that mass determines the periodicity of time in any given coordinate system.  Therefore, in order to properly analyze the transformation to a rotating coordinate system one must specify the mass distribution at rest in the rotating metric.  In other words, the correct analysis of the rotation of an inertial frame must include the specification of the masses that engender the rotation.  The time periodicity defined by the mass distribution at rest in the rotating metric then leads to quantized radii for which the time discontinuity problem is resolved.  For example, a thin, rotating, spherical shell of mass $M$, radius $R$ and angular velocity $\omega_s$ induces a rotation $\omega=\omega_s(4m/3R)$ in its interior \cite{brillcohen}.  From our discussion above, we postulate that the periodicity of coordinate time is determined by the mass at rest in the rotating frame via $\frac{h}{Mc^{2}}$.  Calculating $l_0^{\ast}$ in Eq. (\ref{rotatingquantradii}) we find:
\begin{equation}
\label{sphericalshellradii}
l_0^{\ast}=c\sqrt{\frac{3\hbar R}{4GM^2\omega_s}}.
\end{equation}
For typical laboratory scales $M\sim 10^2$ kg, $R\sim1$ m and $\omega_s\sim 10^1$ s$^{-1}$, which gives $l^{\ast}_0\sim10^{-6}$ m.  

The postulate that coordinate time possesses a periodicity $t^{\ast}$ is consistent with general relativity. General relativity is a local theory; there may be many topologically distinct spacetimes that correspond to a given local metric element $ds^2=g_{\mu\nu}dx^{\mu}dx^{\nu}$ ($\mu,\nu=0,1,2,3$) \cite{reyluminet}.  The above considerations hint at the general relationship that may exist between quantum theory and general relativity.  While general relativity determines the local geometry of space-time, quantum theory may narrow the choice of topology from the topologically distinct space-times that correspond to any given metric.

For an arbitrary line element $ds^2=g_{\mu\nu}dx^{\mu}dx^{\nu}$ the kinematic condition (\ref{quantcond1}) becomes:  
\begin{equation}
\label{quantcond4}
\frac{1}{c}\oint\frac{g_{0i}}{g_{00}}dx^i=nt^{\ast},
\end{equation}
where $i=1,2,3$ and $t^{\ast}$ is the periodicity of coordinate time in the chosen coordinate system.  Note that (\ref{quantcond4}) is not invariant under general coordinate transformations.  Therefore, there is the potential for ambiguity in the choice of $t^{\ast}$ in general solutions of Einstein's equations. 

\section{\label{sec:RotatingMatter}ROTATING DISTRIBUTIONS OF MATTER}
We now apply the above kinematic condition to exact solutions of Einstein's equations for rotating distributions of matter.  First, we consider the Kerr-Newman black hole.  Since the Kerr-Newman metric asymptotically approaches flat space-time the application of the preceding analysis is straightforward.  In quasispheroidal coordinates the Kerr-Newman metric is given by \cite{kerr,newmanetal,carter}:
\begin{eqnarray}
\label{kerrmetric}
ds^2&=&\rho^2d\theta^2-2a\sin^2\theta drd\phi+2drdu+ \nonumber \\
& &\rho^{-2}\left[(r^2+a^2)^2-\Delta a^2\sin^2\theta \right]\sin^2\theta d\phi^2 \nonumber \\
& &-2a\rho^{-2}(2mr-q^2)\sin^2\theta d\phi du -\left[1-\rho^{-2}(2mr-q^2)\right]du^2, 
\end{eqnarray}
where $\rho^2\equiv r^2+a^2\cos^2\theta$, $\Delta\equiv r^2-2mr+a^2+q^2$, $m\equiv\frac{GM}{c^2}$, $q^2\equiv\frac{GQ^2}{c^4}$ and $a\equiv\frac{J}{Mc}$, with $M$ the gravitational mass, $Q$ the charge, and $J$ the angular momentum as observed at large $r$.  Following our analysis above, we postulate that the period of coordinate time in this coordinate system is given by the quantity $\frac{h}{Mc^2}$.  Enforcing Eq. (\ref{quantcond4}) around the axis of rotation, we obtain:
\begin{equation}
\frac{a\rho^{-2}(2mr-q^2)\sin^2\theta}{\left[1-\rho^{-2}(2mr-q^2)\right]}=n\lambda,
\end{equation}
where $\lambda\equiv\frac{\hbar}{Mc}$.  
Consequently, we obtain the following quantized radii:
\begin{equation}
\label{kerrquantradii}
r=m+\frac{am\sin^2\theta}{n\lambda} \pm\left[(m^2-a^2\cos^2\theta-q^2)+\frac{a\sin^2\theta}{n\lambda}(2m^2-q^2)+\frac{a^2m^2\sin^4\theta}{n^2\lambda^2}\right]^{1/2}.
\end{equation}
The characteristic length $\frac{am}{\lambda}$ determines the scale of quantization. An estimate for the sun ($M=2\times10^{30}$ kg and $J\sim10^{41}$ kg m$^2$ s$^{-1}$), gives $\frac{am}{\lambda}\sim10^{79}$.  This suggests $n\sim10^{68}$ for $r\sim 10^{11}$ m, so that an effective continuum of radii exists in our solar system.  Note that for the case $a=0$, the off-diagonal components of the metric vanish and hence no radial quantization results from Eq. (\ref{quantcond4}).   

For $\theta=\pi/2$, Eq. (\ref{kerrquantradii}) gives:
\begin{equation}
r=m+\frac{am}{n\lambda} \pm \left[(m+\frac{am}{n\lambda})^2-q^2(1+\frac{a}{n\lambda})  \right]^{1/2}.
\end{equation}
For $q=0$ this reduces to $r=0$ and $r=2m+(2am/n\lambda)$.  In the uncharged Kerr metric CTCs exist for a small region of negative $r$ in the immediate neighborhood of the singularity $\rho^2=0$ \cite{carter}.  In the charged case, CTC's do not extend beyond the point where $r^2=q^2$ on the positive side of $r$ \cite{carter}.  Therefore, as in the case of the rotating frame of reference, we see that causality violation is avoided in the Kerr-Newman metric.

Next, we calculate the quantized radii for an infinitesimally thin disk of dust rotating uniformly around its symmetry axis \cite{neugebauermeinel1,neugebauermeinel2,neugebauermeinel3}.  The line element in Weyl-Lewis-Papapetrou coordinates is:
\begin{equation}
\label{diskmetric}
ds^2=e^{-2U}\left[e^{2k}\left(d\rho^2+d\zeta^2\right)+\rho^2d\phi^2\right]-e^{2U}\left(dt+ad\phi \right)^2.
\end{equation}
The three metric functions $e^{2U}(\rho,\zeta)$, $e^{2k}(\rho,\zeta)$ and $a(\rho,\zeta)$ depend uniquely on the angular velocity of the disk, $\Omega$, and the relativistic parameter $\mu$:
\begin{equation}
\label{relparameter}
\mu=\frac{2\Omega^2\rho_0^2e^{-2V_0}}{c^2},
\end{equation}
where $\rho_{0}$ is the coordinate radius of the disk and $V_0\equiv U(\rho=0,\zeta=0)$ is the ``surface potential''.  The metric functions are calculated from the complex Ernst equation along with the appropriate boundary conditions.

Noting that metric (\ref{diskmetric}) is asymptotically flat, we postulate that the period of coordinate time in this coordinate system is given by the quantity $\frac{h}{Mc^2}$, where $M$ is the total gravitational mass determined uniquely by $\mu$ and $\rho_0$.  Enforcing condition (\ref{quantcond4}) around the axis of rotation we obtain:
\begin{equation}
\label{quantconddisk1}
a(\rho,\zeta)=n\lambda,
\end{equation}
where $\lambda\equiv\frac{\hbar}{Mc}$.  We consider the Newtonian limit $\mu= 2\Omega^2\rho_0^2/c^2\ll 1$.  At $\mu=0$ the metric functions $e^{2U}(\rho,\zeta)$ and $a(\rho,\zeta)$ may be expanded in a power series \cite{neugebauermeinel3}:
\begin{eqnarray}
\label{sums}
e^{2U}(\rho,\zeta)&=&1 + \sum_{n=1}^\infty f_{2n-1}\mu^n \nonumber \\
a(\rho,\zeta)&=&\sum_{n=1}^\infty a_{2n}\mu^{(2n+1)/2},
\end{eqnarray}
where the coefficients $f_{2n-1}$ and $a_{2n}$ are elementary functions and are calculated in Ref. \cite{petroffmeinel}.  Consequently, Eq. (\ref{quantconddisk1}) gives (to lowest order in $\mu$):
\begin{equation}
\label{quantconditiondisk2}
a_2\mu^{3/2}=n\lambda.
\end{equation}
Therefore, we obtain the following quantized radii in the plane of the disk ($\zeta=0$):
\begin{equation}
\label{radiidisk}
\left(\frac{\rho}{\rho_0}\right)^2=\frac{2}{3}\pm\frac{2}{3}\left[1-n\sqrt{\frac{2}{\mu^3}}\frac{\lambda}{\rho_0}\right]^{1/2}.
\end{equation} 
Galaxies or galaxy clusters can be modeled to a first approximation as rotating disks of dust.  For typical galactic scales, $\frac{\lambda}{\rho_0}\ll \mu^{3/2}$, therefore Eq. (\ref{radiidisk}) predicts an effective continuum of preferred radii.   

From the line element (\ref{diskmetric}) we obtain the condition for CTCs:
\begin{equation}
\label{diskctc1}
g_{\phi\phi}=\rho^2e^{-2U}-e^{2U}a^2<0.
\end{equation}
With the power series expansions (\ref{sums}) this becomes (to lowest order in $\mu$):
\begin{equation}
\label{diskctc2}
3\left(\frac{\rho}{\rho_0}\right)^3-4\left(\frac{\rho}{\rho_0}\right) >4\left(\frac{2}{\mu^3}\right)^{1/2},
\end{equation}
for $\zeta=0$.  In the Newtonian limit $\mu\rightarrow 0$, so that CTCs occur for $\rho/\rho_0 \rightarrow \infty$.  Since the quantized radii (\ref{radiidisk}) are bounded by the condition $\left(\frac{\rho}{\rho_0}\right)^2<\frac{4}{3}$ for all $n$, causality violation is avoided in this metric as well.  

\section{\label{sec:discussion}DISCUSSION}
In conclusion, a kinematic condition is proposed that resolves the problem of the time discontinuity in rotating space-times.  This condition predicts radial quantization in both the quantum and macroscopic domains.  For the cases considered above, the kinematic condition predicts radii that do not lie in the regions that possess CTC's.  These quantized radii suggest a new manifestation of Hawking's chronology protection conjecture \cite{Hawking}.  Moreover, such radial quantization may eliminate the need for the cosmic censorship hypothesis \cite{Penrose}.  Further work is needed to examine this condition in the context of other solutions of Einstein's equations that possess CTCs (i.e., the rotating G\"{o}del universe, cosmic strings, etc.) to see if causality violation is prevented in these cases as well.  However, these metrics are not in general asymptotically Minkowskian; therefore, the definition of $t^{\ast}$ is not well-defined. 

We have postulated that the presence of mass affects not only the local properties of space-time, but also its topological structure.  Specifically, we proposed that the coordinate $t$ usually identified with physical time is periodic under circumnavigation in the rest frame of the mass.  We can recover the time of our physical experience in the following manner.  Without loss of generality we may align the direction of circumnavigation with the $\hat{\phi}$ direction of a cylindrical coordinate sytem.  Hence, the usual identification of points
\begin{equation}
\label{identifypoints1}
\{t,r,\phi,z\} =  \{t,r,\phi+2\pi,z\} 
\end{equation}
must be replaced in the presence of a mass $M$ by 
\begin{equation}
\{t,r,\phi,z\} =  \{t+2\pi  \frac{\hbar}{Mc^2},r,\phi+2\pi,z\}, 
\end{equation}
so that $t$ possesses topology $S^1$.  Consider the following transformation:
\begin{equation}
\tilde{t} = t - \frac{\hbar}{Mc^2}\phi,
\end{equation} 
where $\tilde{t}$ is defined to be $R^1$.  This permits the following identification of points in the manifold:
\begin{equation}
\{\tilde{t},r,\phi,z\} =  \{\tilde{t},r,\phi+2\pi,z\}.
\end{equation}
The difference between the time $\tilde{t}$ and coordinate time $t$ may be understood by introducing Killing vector fields.  Flat space-time (i.e., the asymptotic space-time of a mass $M$) is assumed to possess the time-translational and rotational Killing fields $\xi^{\mu}=(\partial/\partial t)^{\mu}$ and $\psi^{\mu}=(\partial/ \partial \phi)^{\mu}$ (in addition to $(\partial/ \partial \theta)^{\mu}$ if there is no spin), with $t$ unbounded and $\phi$ periodic, i.e., $\phi = \phi+2\pi n$.  This periodicity results from the identification of points along orbits of the rotational Killing field $\psi^{\mu}$.  However, the considerations above suggest that the asymptotic flat space-time of a mass $M$ should be defined rather with the identification of points along the helical orbits of $\tilde{\psi}^{\mu}$ defined by
\begin{equation}
\tilde{\psi}^{\mu} = \psi^{\mu}+\frac{\hbar}{Mc^2}\xi^{\mu}.
\end{equation}
Thus, we conclude that the physical time of our experience is not the coordinate $t$ with topology $S^1$ which is constant along orbits of $\psi^{\mu}$ but it is the coordinate $\tilde{t}$ with topology $R^1$ which is constant along orbits of $\tilde{\psi}^{\mu}$.  However, since  $\hbar/c^2 \sim 10^{-50}$ the coordinate constant along orbits of $\psi^{\mu}$, i.e. $t$, may be identified with the time of our physical experience to first approximation.  These considerations force us to reconsider the asymptotic nature of the well-known solutions to Einstein's equations with localized mass distributions. 

The kinematic condition proposed above suggests that the dragging of inertial frames is subject to the well-known quantum condition for a massive particle with orbital angular momentum (see Eq. (\ref{stationarystates})).  This condition may be written:   
\begin{equation}
\frac{Mg_{\phi\phi}\Omega}{\sqrt{g_{tt}}}=n\hbar,
\end{equation}
where $M$ is the gravitational mass that causes the frame-dragging and $\Omega=\left|\frac{g_{t\phi}}{g_{\phi\phi}}\right|$ is the frame-dragging frequency.  In other words, the wave-like nature associated with the motion of an inertial frame that is identified with the rest frame of a massive particle must be generalized to apply to the inertial frames dragged along by a rotating mass as well.  Hence, in analogy with the stationary states of an electron, one concludes that frame dragging can only occur at radii that satisfy the kinematic condition and is consequently forbidden for radii that do not satisfy the kinematic condition.  

As is well known, quantized planetary orbits exist in the solar system.  The quantized radii predicted in the cases above do not correspond with these observed values.  However, this does not preclude the possibility that the observed planetary orbits are remnants of such radial quantization from an earlier stage of the solar system evolution.  For example, the above metrics may not accurately depict the early nebula of dust and gas, which when flattened into a disk possessed a bulge at its center.  Further work is needed to determine the quantized radii in more accurate models.
 
\bibliographystyle{unsrt}
\bibliography{qr}

\begin{thebibliography}{10}

\bibitem{relrotbook}
G.~Rizzi and M.~L. Ruggiero, editors.
\newblock {\em Relativity in Rotating Frames}.
\newblock Kluwer Academic, Dordrecht, 2004.

\bibitem{lobocrawford}
F.~Lobo and P.~Crawford.
\newblock Time, closed timelike curves and causality.
\newblock In R.~Buccheri, M.~Saniga, and W.~M. Stuckey, editors, {\em The
  Nature of Time: Geometry, Physics and Perception, NATO Science Series II.
  Mathematics, Physics and Chemistry}, volume~95, page 289. Kluwer Academic,
  Dordrecht, 2003.

\bibitem{dieks}
D.~Dieks.
\newblock Space, time and coordinates in a rotating world.
\newblock In G.~Rizzi and M.~L. Ruggiero, editors, {\em Relativity in Rotating
  Frames}, page~29. Kluwer Academic, Dordrecht, 2004.

\bibitem{weber}
T.~Weber.
\newblock Elementary considerations of the time and geometry of rotating
  reference frames.
\newblock In G.~Rizzi and M.~L. Ruggiero, editors, {\em Relativity in Rotating
  Frames}, page 139. Kluwer Academic, Dordrecht, 2004.

\bibitem{anandan}
J.~Anandan.
\newblock {\em Phys. Rev. D}, 24:338, 1981.

\bibitem{bergiaguidone}
S.~Bergia and M.~Guidone.
\newblock {\em Found. Phys. Lett.}, 11:549, 1998.

\bibitem{cranoretal}
M.~B. Cranor, E.~M. Heider, and R.~H. Price.
\newblock {\em Am. J. Phys.}, 68:1016, 2000.

\bibitem{rizzitartaglia}
G.~Rizzi and A.~Tartaglia.
\newblock {\em Found. Phys.}, 28:1663, 1998.

\bibitem{klauber2}
R.~Klauber.
\newblock Toward a consistent theory of relativistic rotation.
\newblock In G.~Rizzi and M.~L. Ruggiero, editors, {\em Relativity in Rotating
  Frames}, page 103. Kluwer Academic, Dordrecht, 2004.

\bibitem{selleri2}
F.~Selleri.
\newblock Sagnac effect: end of the mystery.
\newblock In G.~Rizzi and M.~L. Ruggiero, editors, {\em Relativity in Rotating
  Frames}, page~57. Kluwer Academic, Dordrecht, 2004.

\bibitem{nahin}
P.~J. Nahin.
\newblock {\em Time Machines: Time Travel in Physics, Metaphysics and Science
  Fiction}.
\newblock Springer-Verlag and AIP Press, New York, 1999.

\bibitem{vanstockum}
W.~J. van Stockum.
\newblock {\em Proc. Roy. Soc. Edinburgh}, 57:135, 1937.

\bibitem{tipler}
F.~Tipler.
\newblock {\em Phys. Rev. D}, 9:2203, 1974.

\bibitem{godel}
K.~Godel.
\newblock {\em Rev. Mod. Phys.}, 21:447, 1949.

\bibitem{carter}
B.~Carter.
\newblock {\em Phys. Rev.}, 174:1559, 1968.

\bibitem{deseretal}
S.~Deser, R.~Jackiw, and G.~'t~Hooft.
\newblock {\em Ann. Phys. (NY)}, 152:220, 1984.

\bibitem{gott}
J.~Gott.
\newblock {\em Phys. Rev. Lett.}, 66:1126, 1991.

\bibitem{wald}
R.~M. Wald.
\newblock {\em Phys. Rev. Lett.}, 26:1653, 1971.

\bibitem{mallett}
R.~L. Mallett.
\newblock {\em Found. Phys.}, 33:1307, 2003.

\bibitem{bonnor2}
W.~B. Bonnor.
\newblock {\em Class. Quantum Grav.}, 18:1381, 2001.

\bibitem{bonnor3}
W.~B. Bonnor.
\newblock {\em Phys. Letters A}, 284:81, 2001.

\bibitem{bonnorward}
W.~B. Bonnor and J.~P. Ward.
\newblock {\em Commun. Math. Phys.}, 34:123, 1973.

\bibitem{bicakpravda}
J.~Bicak and V.~Pravda.
\newblock {\em Phys. Rev.}, 60:044004, 1999.

\bibitem{bonnor}
W.~B. Bonnor.
\newblock {\em Int. J. Mod. Phys.}, D12:1705, 2003.

\bibitem{Hawking}
S.~Hawking.
\newblock {\em Phys. Rev. D}, 46:603, 1992.

\bibitem{kimthorne}
S.-W. Kim and K.~S. Thorne.
\newblock {\em Phys. Rev. D}, 43:3929, 1991.

\bibitem{landaulifshitz}
L.~Landau and E.~Lifshitz.
\newblock {\em The Classical Theory of Fields}.
\newblock Pergamon Press, Oxford, 1971.

\bibitem{debroglie}
L.~de~Broglie.
\newblock {\em C. R. Acad. Sci.}, 177:507, 1923.

\bibitem{brillcohen}
D.~R. Brill and J.~M. Cohen.
\newblock {\em Phys. Rev.}, 143:1011, 1966.

\bibitem{reyluminet}
M.~Lachi$\grave{\textrm{e}}$ze-Rey and J.~P. Luminet.
\newblock {\em Phys. Rep.}, 254:135, 1995.

\bibitem{kerr}
R.~P. Kerr.
\newblock {\em Phys. Rev. Lett.}, 11:238, 1963.

\bibitem{newmanetal}
E.~T. Newman, E.~Couch, R.~Chinnapared, A.~Exton, A.~Prakash, and R.~Torrence.
\newblock {\em J. Math. Phys.}, 6:918, 1965.

\bibitem{neugebauermeinel1}
G.~Neugebauer and R.~Meinel.
\newblock {\em Astrophys. J.}, 414:L97, 1993.

\bibitem{neugebauermeinel2}
G.~Neugebauer and R.~Meinel.
\newblock {\em Phys. Rev. Lett.}, 73:2166, 1994.

\bibitem{neugebauermeinel3}
G.~Neugebauer and R.~Meinel.
\newblock {\em Phys. Rev. Lett.}, 75:3046, 1995.

\bibitem{petroffmeinel}
D.~Petroff and R.~Meinel.
\newblock {\em Phys. Rev. D}, 63:064012, 2001.

\bibitem{Penrose}
R.~Penrose.
\newblock {\em Revistas del Nuovo Cimento}, 1:252, 1969.

\end{thebibliography}

\end{document}